\documentclass[12pt,preprint]{aastex}
\usepackage{epsfig}
\def\ls{{_<\atop^{\sim}}}
\def\gs{{_>\atop^{\sim}}}

\begin{document}

\title{Chandra Discovery of a Tree in the X-ray Forest towards
PKS~2155-304: the Local Filament?}

\author{Fabrizio Nicastro$^1$, Andreas Zezas$^1$, Jeremy
Drake$^1$, Martin Elvis$^1$, Fabrizio Fiore$^2$, Antonella
Fruscione$^1$, Massimo Marengo$^1$, Smita Mathur$^3$, Stefano Bianchi$^4$}

\affil {$^1$ Harvard-Smithsonian Center for Astrophysics\\ 
60 Garden St, Cambridge MA 02138}

\affil {$^2$ Osservatorio Astronomico di Roma, Via Frascati 33, 
Monteporzio-Catone (RM), I-00040 Italy}

\affil {$^3$ Astronomy Department, The Ohio State University, \\
43210, Columbus, OH, USA}

\affil {$^4$ Dipartimento di Fisica, Universit\'a degli Studi
Roma Tre, \\ Via della Vasca Navale 84, I-00146 Roma, Italy}

\author{\tt version: 28 December, 2001}

\begin {abstract}
We present the first X-ray detection of resonant absorption from
warm/hot local gas either in our Galaxy, or in the
Intergalactic space surrounding our Galaxy, along the line of
sight toward the blazar PKS~2155-304.  The {\em Chandra}
HRCS-LETG spectrum of this z=0.116 source clearly shows, at $\ge 5\sigma$ 
level, unresolved (FWHM $\le 800$ km $s^{-1}$ at a 2$\sigma$ confidence level)
OVII$_{K\alpha}$ and NeIX$_{K\alpha}$ resonant absorption lines
at $21.603^{+0.014}_{-0.024}$ \AA\ and $13.448^{+0.022}_{-0.024}$
\AA\ (i.e. $cz = (14^{+190}_{-330})$ km s$^{-1}$ in the rest
frame, from the OVII$_{K\alpha}$ line).
OVIII$_{K\alpha}$ and OVII$_{K\beta}$ from the same system are
also detected at a lower significance level (i.e. $\sim 3\sigma$), while 
upper limits are set on OVIII$_{K\beta}$, NeX$_{K\alpha}$ and NeIX$_{K\beta}$.
The FUSE spectrum of this source shows complex OVI$_{2s
\rightarrow 2p}$ absorption at the same redshift as the X-ray
system, made by at least two components: one relatively narrow
(FWHM $=106 \pm 9$ km s$^{-1}$) and slightly redshifted ($cz = 36
\pm 6$ km s$^{-1}$), and one broader (FWHM $= 158 \pm 26$ km
s$^{-1}$) and blueshifted ($cz = -135 \pm 14$ km s$^{-1}$).  We
demonstrate that the physical states of the UV and X-ray
absorbers are hard to reconcile with a single, purely
collisionally ionized, equilibrium plasma.  We propose, instead,
that the X-ray and, at least the broader and blueshifted UV
absorber are produced in a low density intergalactic plasma,
collapsing towards our Galaxy, consistent with the predictions of a 
Warm-Hot Intergalactic Medium (WHIM) from numerical simulations. 
We find that any reasonable
solution requires overabundance of Ne compared to O by a factor
of $\sim 2$, with respect to the solar value. We propose several
scenarios to account for this observation.
\end{abstract}

\newpage

\section{\bf Introduction}

The majority of the total baryonic matter in the local ($z\ls 1$)
Universe is predicted to be concentrated in highly ionized
gas. Structures that are already virialized contain warm
($10^5-10^6$ K) or hot (10$^7$ K) gas [the dense interstellar
medium (ISM) of galaxies, and the intracluster medium (ICM) of
clusters of galaxies].  The greater amount of baryonic matter is
predicted to lie in, as yet unvirialized, matter in the form of a
tenuous warm-hot intergalactic medium (WHIM, Hellsten et al.,
1998).  The detection and study of these components is needed for
the proper understanding of large and small scale structures in
the Universe.  However, while X-ray emission from the virialized
density peaks of the ICM and ISM has been detected and
intensively studied in X-rays, the predicted highly ionized gas
in the WHIM has been poorly studied so far, due to instrumental
limitations. The low density of the WHIM leads to low emissivity,
so that studies of the WHIM in emission are a formidable
challenge. However, absorption depends only on the total column
density of the medium, not on density, and background light
sources in the form of quasars (Aldcroft et al., 1994) and
gamma-ray bursts (Fiore et al. 2000) are readily available.  A
few high ionization transitions, notably OVI $\lambda$1031.93,
lie in the UV, but the most prominent ions (CVI, OVII, OVIII,
NeIX) have their strongest transitions in the soft X-ray band
(10-40 \AA) for a wide range of temperatures ($10^5-10^{6.5}$ K,
Nicastro et al., 1999), which should give rise to an ``X-ray
Forest'' of absorption lines (e.g. Perna \& Loeb, 1998, Fang \&
Canizares, 2000).  The advent of high resolution ($R\sim$1000)
soft X-ray spectroscopy with {\em Chandra} and XMM-{\em Newton}
allows sensitive studies of the WHIM possible.


Interstellar OVI was first detected in the UV with the {\em
Copernicus} satellite (Jenkins, 1978a, 1978b, and York, 1977),
but only recently has data from the {\em Far Ultraviolet
Spectroscopic Explorer} (FUSE) satellite shown the presence of
extragalactic OVI intervening absorpion (e.g.  Tripp et al.,
2001, Sembach et al., 2000).


In this paper we present the first X-ray detection of highly
ionized absorption along the line of sight towards the blazar
PKS~2155-304, with the {\em Chandra} HRCS-LETG ({\em High
Resolution Camera Spectrometer}-{\em Low Energy Transmission
Grating}: Brinkman et al., 2000).  We argue that the physical and
dynamical conditions implied by these lines require an
extragalactic, low density, origin. Hence these lines are the
first detection of the X-ray Forest.


\section{Spectral Fitting}

The bright z=0.116 blazar PKS~2155-304 (Falomo, Pesce, \& Treves,
1993) is a calibration source for both the high energy and low
energy resolution gratings of {\em Chandra}, and has been
observed several times with all the possible grating-detector
configurations. The longest of these observation lasted $\sim
64.7$ ks, for a total net good-time exposure of 62.7 ks, and was
performed with the HRCS-LETG configuration on 1999 December
25. In the public {\em Chandra} data archive
\footnote{http://asc.harvard.edu/cda/}
we found two more HRCS-LETG observations of this source,
performed on 2000 May 31, and 2001 April 6, with net exposures of
25.8 ks and 26.6 ks respectively.

We retrieved the primary and secondary data products (Fabbiano et
al., 2001, in prep.) of these three {\em Chandra} HRCS-LETG
observations of PKS~2155-304, and reprocessed their event files
with the {\em CIAO} (Elvis et al., 2001, in prep.) software
(v. 2.1.3) using the most up-to-date calibration files as of 2001
August (CALDB v. 6.2) to extract source and background spectra
and corresponding 1st-order Anciliary Response Files (ARFs) and
Redistribution Matrices (RMFs) according to the on-line data
analysis ``threads'' provided by the {\em Chandra X-ray Center}
(CXC)
\footnote{http://asc.harvard.edu/ciao/documents\_threads.html}
. 

During the long 1999 December observation the source was at the
particularly high flux level of 21 mCrab, with a 0.3-6 keV flux
of $F_{2-40 \AA} = 4.2 \times 10^{-10}$ erg s$^{-1}$
cm$^{-2}$. In both the subsequent HRCS-LETG observations,
instead, the flux level of PKS~2155-304 was more normal at about
6 mCrab in the 0.3-6 keV band.  However, despite the large change
in flux, the spectral shape of the underlying 0.3-6 keV continuum
did not change significantly between or during the three
observations (using simple power law continuum models, we
measured variations in the spectral slope to be less than $\Delta
\Gamma = 0.1$ at 90 \% confidence level).  To increase the signal
to noise ratio of our data, we then coadded the three HRCS-LETG
spectra and averaged the corresponding ARFs, weighting them by
the respective exposure times. The result (Figure 1a) is one of 
the highest quality {\em Chandra} grating spectra so far for an
extragalactic source, with some 600 counts, or a signal to noise
ratio of about 25, per resolution element (R$\sim 450$, FWHM $\simeq$ 
660 km s$^{-1}$ at 20 \AA) in the continuum, over the 
entire 10-25 \AA\ wavelength ($\sim 0.5-1.2$ keV) range. 

\medskip 
To ensure a self-consistent analysis, and because of the $\sim
30$ times higher spectral resolution provided, we also retrieved
the FUSE (Moos et al., 2000) data of PKS~2155-304 from the FUSE
public archive, extracted the fluxed spectrum, and analyzed the
wavelength range around OVI (1027-1035 \AA) and OI (1037-1040
\AA) to obtain independent estimates of the positions, widths and
equivalent widths of the local OVI$_{2s\rightarrow
2p}(\lambda=1031.9261)$ and OI$_{3s\rightarrow
3p}$($\lambda=1039.23$) components, recently discovered and
published by Savage et al. (2000) and Sembach et al. (2000).  The
FUSE observation was taken with the LWRS aperture
($30"\times30"$) with a total exposure time of 38.6 ks, and
covers the entire 980-1180 \AA\ range. We used data from the LiF
mirror and the A1 detector segment, which have optimal efficiency
in the waveband around the OVI line.  The data were calibrated
and cleaned following the procedures presented in the FUSE Data
Analysis Cookbook (v. 1.0, Sankrit et al., 2001). The resolution
of the final data is $\sim20$ km s$^{-1}$. 

We performed simultaneous spectral fitting of both the fluxed
FUSE data and the responses-folded {\em Chandra} HRCS-LETG data,
using the {\em Sherpa} (Siemiginowska et al., 2001, in prep.)
package for spectral fitting in {\em CIAO}. In the following two
sections we summarize our results.

\subsection{X-ray Spectral Fitting}

For sources of a few mCrab the inherent order confusion problem
of the {\em Chandra} HRCS-LETG (Brinkman et al., 2000) is
important only at $\lambda \gs 25$ \AA. Furthermore, narrow
features, such as absorption lines, are not in any case seriously
affected.  So we limited our spectral analysis to the shorter
wavelengths of 11-24 \AA\ ($\sim 0.5-1.2$ keV) range, allowing us
to use only the 1st-order RMFs and ARFs, to fit our spectra.
Errors throughout the paper are quoted at a 90 \% confidence
level, for 1 interesting parameter.

\medskip
To model the 11-24 \AA\ portions of the continuum of
PKS~2155-304, we used a power law attenuated by an equivalent
hydrogen column density of $N_H = 1.65 \times 10^{20}$ cm$^{-2}$
neutral gas to account for the measured Galactic absorption from
cold gas along the line of sight to this source (Dickey \&
Lockman, 1990)
\footnote{Due to the high degree of saturation of the Galactic disk $N_H$, 
the local damped Ly$\alpha$ absorption line in the HST-STIS
spectrum of PKS~2155-304 cannot provide a better estimate of the
Galactic column of neutral H along this line of sight.}
. The model used for the neutral absorber includes only
photoelectric edges by neutral elements, and so does not account
for resonant absorption lines from the same elements.  We
obtained a statistically acceptable fit with a power law photon
index of $\Gamma = 2.42 \pm 0.02$, consistent previous results 
(e.g. Kataoka et al., 2000). 

The residuals to this fit clearly show (at a significance level
greater than 8 $\sigma$) two strong and unresolved (see \S 3.1)
features in absorption (Fig. 1b,d), at wavelengths consistent
with the OVII$_{K\alpha}$ and the NeIX$_{K\alpha}$ resonant lines
at zero velocity, plus a third intense
OI$_{1s\rightarrow 2p}$ resonant absorption line from gaseous OI
(Fig. 1c).  Negative residuals are also present at the energy of
the instrumental resonant molecular OI absorption line (Fig. 1c).
The residuals around the expected OVIII$_{K\alpha}$ and the
OVII$_{K\beta}$ rest-frame wavelengths also show absorption
features at a lower ($\sim 3 \sigma$) significance level
(Fig. 1e, see \S 3.1).  We added six negative gaussians to our
continuum model, and refitted the data, leaving the positions and
the total flux of all the gaussians free to vary independently,
but linking all the widths (in units of km s$^{-1}$) of the high
ionization resonance lines from O and Ne.
%

\begin{figure}
\plotone{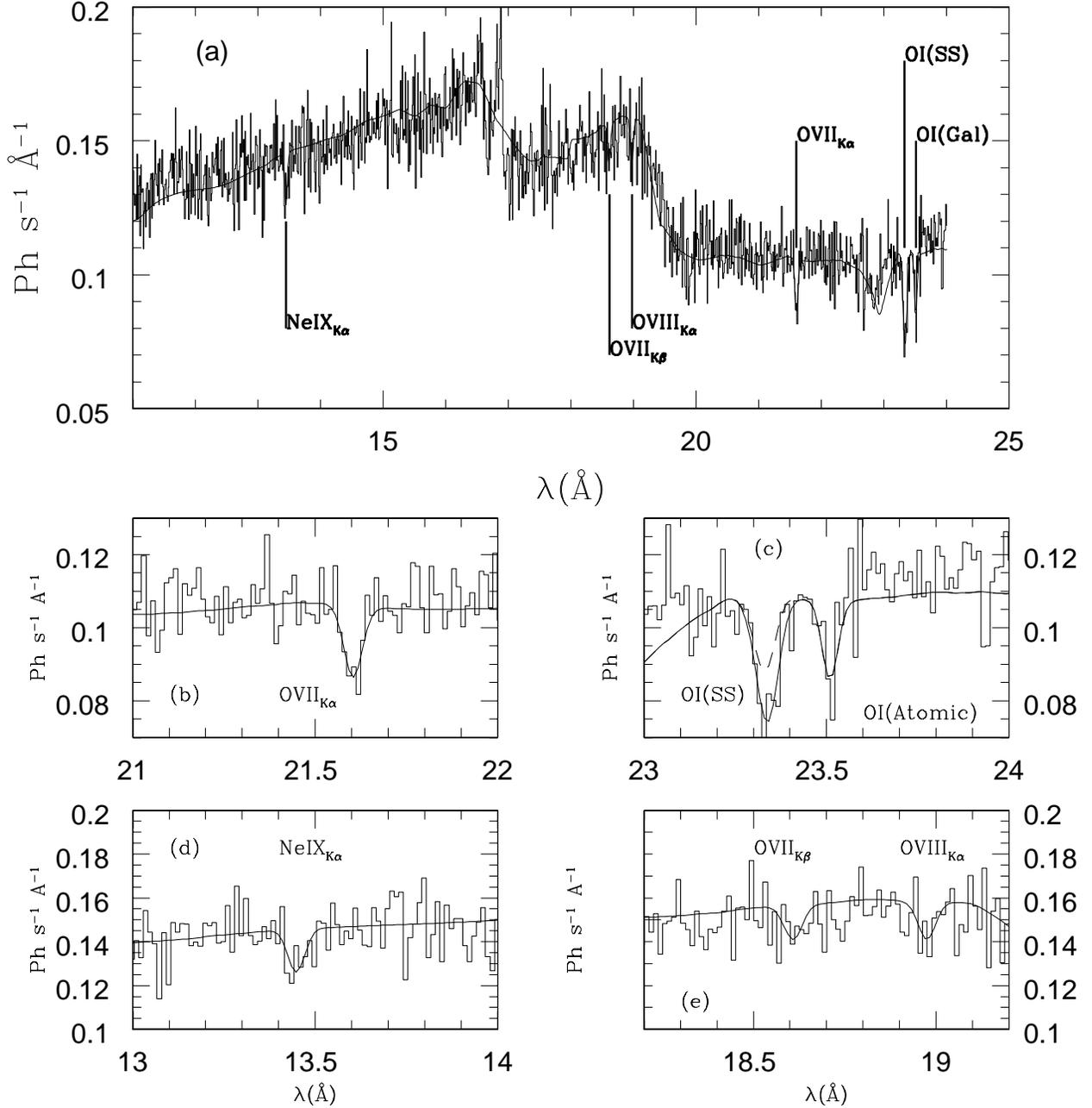} 
\vspace{0in}\caption[h]{\footnotesize Top: (a) X-ray 11-24 \AA\
unbinned ($\Delta \lambda = 12.5$ m\AA) HRCS-LETG spectrum of
PKS~2155-304. The folded best fitting model (power-law plus
Galactic absorption, plus 6 absorption lines) is
shown as a solid line.  The strongest absorption lines are
identified. 
Note that the effective area of the HRCS-LETG has not been
corrected for in this plot, in order to show the strong, but
large scale, instrumental features.
The fitted power-law follows these peaks and dips closely.
Bottom: blow-ups of four portions of the spectrum in
Fig. 1a: (b) OVII$_{K\alpha}$ (top left); (c) atomic and
solid-state (i.e. molecular) OI (top right); (d) NeIX$_{K\alpha}$
(bottom left); (e) OVII$_{K\beta}$ and OVIII$_{K\alpha}$ (bottom
right). Vertical lines indicate the rest frame wavelengths of the 
transitions. The solid-state OI line in Fig. 1c is an effective area
feature (dashed line).}
\end{figure}

%
The best fitting line parameters values are listed in Table 1, along 
with the derived ion column densities for all the atomic transitions. 
We also include in table~1 2$\sigma$ upper limits on the
equivalent widths of the two lines of NeX$_{K\alpha}$ and
NeIX$_{K\beta}$, which we derived by adding two more gaussians to
our model, freezing their positions to the rest-frame wavelengths
of these transitions and calculating the errors on the lines'
normalizations.
%

\begin{table}
\footnotesize
\begin{center}
\caption{\bf \small Best fitting UV and X-ray absorption line
parameters} 
\vspace{0.4truecm}
\begin{tabular}{|c|cccccc|}
\hline
& Line Id. & $\lambda$ & $cz$          & FWHM          &EW(abs)$^a$ & 
N$_{ion}$ \\
& & (\AA)     & (km s$^{-1}$) & (km s$^{-1}$) & (eV) & ($10^{15}$ cm$^{-2}$) \\
\hline
{\em FUSE} & & & & & & \\
& OVI$_{2s\rightarrow 2p}^{Narrow}$ & $1032.05 \pm 0.02$ & $36 \pm 6$ 
& $106 \pm 9$ & $(2.1 \pm 0.2) \times 10^{-3}$ & $0.14 \pm 0.10$ \\
& OVI$_{2s\rightarrow 2p}^{Broad}$ & $1031.46 \pm 0.05$ & $-135 \pm 14$ 
& $158 \pm 26$ & $(1.6 \pm 0.4) \times 10^{-3}$ & $0.11 \pm 0.30$ \\
\hline
{\em Chandra} & & & & & & \\
& OVII$_{K\alpha}$ & $21.603^{+0.014}_{-0.024}$ & $14^{+194}_{-334}$ & 
$< 800$ & $(0.31^{+0.20}_{-0.16})$ & $4.0^{+2.6}_{-2.1}$ \\
& OVIII$_{K\alpha}$ & $18.973_{-0.036}^{+0.022}$ & $63_{-570}^{+348}$ & 
$\equiv$ FWHM(OVII$_{K\alpha})$ & $(0.24^{+0.20}_{-0.18})$ & 
$5.2^{+4.3}_{-3.9}$ \\
& OVII$_{K\beta}$ & $18.605_{-0.088}^{+0.056}$ & $-380^{+900}_{-1400}$ & 
$\equiv$ FWHM(OVII$_{K\alpha})$ & $(0.22^{+0.26}_{-0.20})$ & 
$13^{+15}_{-12}$ \\
& NeIX$_{K\alpha}$ & $13.448^{+0.022}_{-0.024}$ & $20_{-536}^{+490}$ & 
$\equiv$ FWHM(OVII$_{K\alpha})$ & $(0.53^{+0.36}_{-0.34})$ & 
$6.7^{+4.6}_{-4.3}$ \\
& NeX$_{K\alpha}$ & 12.134$^b$ & 0$^b$ & $\equiv$ FWHM(OVII$_{K\alpha})$ & 
$< 0.32$ & $< 5.2$ \\
& NeIX$_{K\beta}$ & 11.5467$^b$ & 0$^b$ & $\equiv$ FWHM(OVII$_{K\alpha})$ 
& $< 0.24$ & $< 20$ \\
& & & & & & \\
& OI($\lambda=23.489^c$) & $23.509^{+0.008}_{-0.018}$ & 
$-140^{+114}_{-230}$ & $< 170$ & -($0.25^{+0.12}_{-0.12}$) & $21 \pm 10$ \\
& OI(molecular) & $23.341^{+0.026}_{-0.018}$ & NA & $< 505$ & 
-($0.30^{+0.22}_{-0.18}$) & \\
\hline
\end{tabular}
\end{center}
$^a$ Absorption line EW are formally negative; here we drop the sign.
$^b$ Frozen. 
$^c$ Krause, 1994. 
\end{table}
\normalsize
%

The strong solid-state instrumental OI line at $\sim 23.3$ \AA\
(Fig. 1a,c), is only partly accounted for by the current
effective area model (Fig. 1c, dashed curve; see also Table
1) and is easily separated by the HRCS-LETG from the gas phase
OI$_{1s\rightarrow 2p}$ resonant transition at $\lambda(OI) =
23.489$ \AA\ (Krause, 2001), which is also clearly visible in our
data (Fig. 1c), and will be object of study of a forthcoming
paper.  Here we stress only how the intervening Galactic cold
absorber is clearly distinguished from the highly ionized
intervening medium seen through the He-like and H-like resonant
lines from O and Ne (Table 1; see also \S 3.1 for differences in
the dynamics).

\subsection{UV Spectral Fitting}

We also used {\em Sherpa} to fit the 1027-1035 \AA\ portion of
the FUSE spectrum of PKS~2155-304 around the strongest line of
the OVI$_{2s\rightarrow 2p}$ doublet, at $\lambda = 1031.9261$
\AA.  Residuals after a fit with a power law continuum model,
show resolved OVI$_{2s\rightarrow 2p}$ absorption in these data,
with at least two components present (Figure 2).  We added two
negative gaussians to our best fit continuum model.  The fit
shows that one line is relatively narrow (FWHM $=106 \pm 9$ km
s$^{-1}$) and slightly redshifted ($cz = 36 \pm 6$ km s$^{-1}$ in the 
rest frame), and the other is broader (FWHM $= 158 \pm 26$
km s$^{-1}$) and blueshifted ($cz = -135 \pm 14$ km s$^{-1}$; see
Table 1).  These values are consistent with those reported by
Savage et al.  (2000) and Sembach et al. (2000).

%
\begin{figure}
\plotone{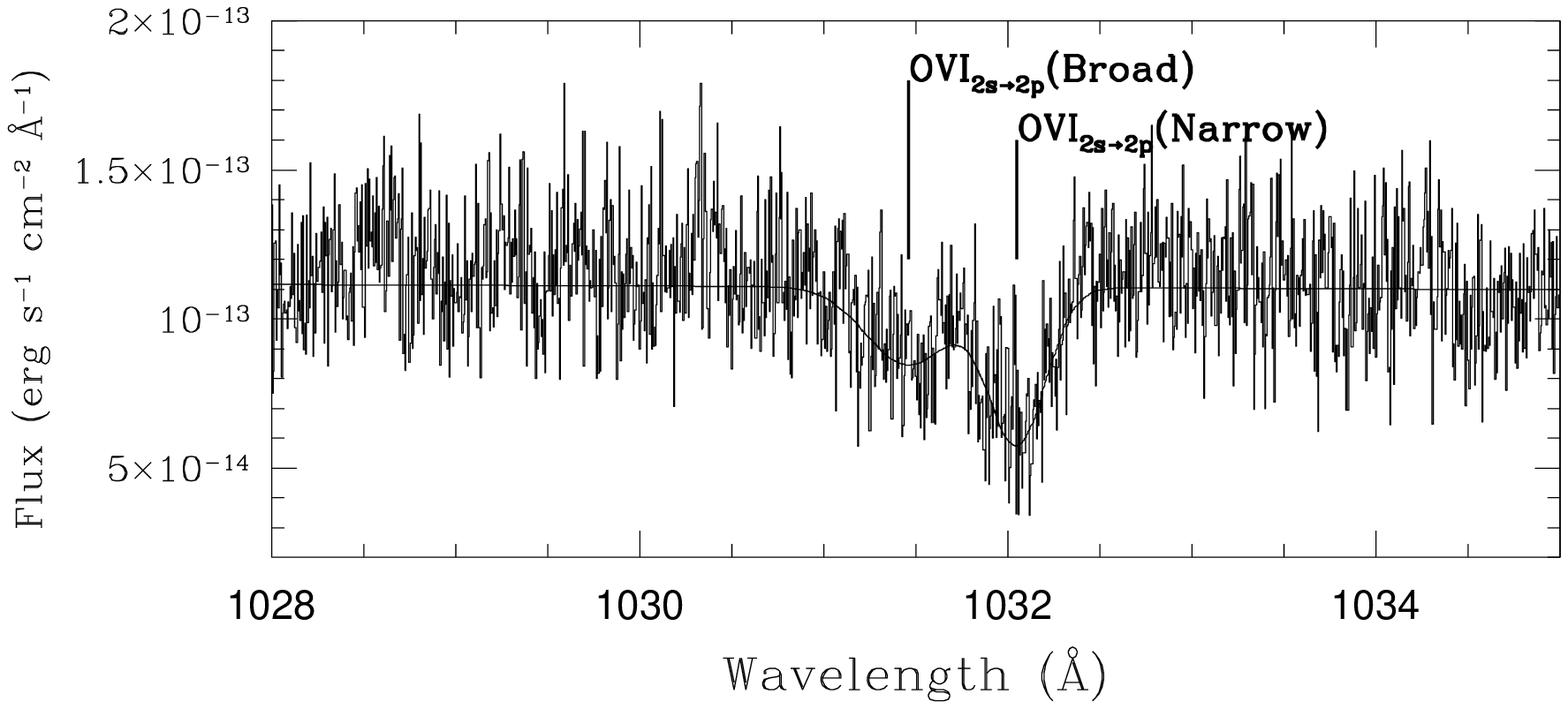} 
\vspace{0in}\caption[h]{\footnotesize Ultraviolet 1027-1035 \AA\
fluxed FUSE spectrum of PKS~2155-304, and best fitting model.}
\end{figure}
%

{\bf We looked for Ly$\alpha$ absorption associated with these
two components, but verified that they both would fall well
within the Voigt wings of the damped Ly$\alpha$ due to cold
absorption in our own Galaxy, and are so not visible. }

The FUSE spectrum of PKS~2155-304 also shows the presence of
strong OI$_{3s\rightarrow 3p}$($\lambda=1039.23$) absorption (see
Figure 1 in Sembach et al., 2000). For this line we measure
$\lambda = 1039.19 \pm 0.02$ ($cz = -11 \pm 6$ km s$^{-1}$) and
FWHM of $47 \pm 5$ km s$^{-1}$, much narrower than the width of
either of the OVI$_{3s\rightarrow 3p}$ components.

\section{Analysis of Absorption Lines}

\subsection{Gas Dynamics}

%
\begin{figure}
\plotone{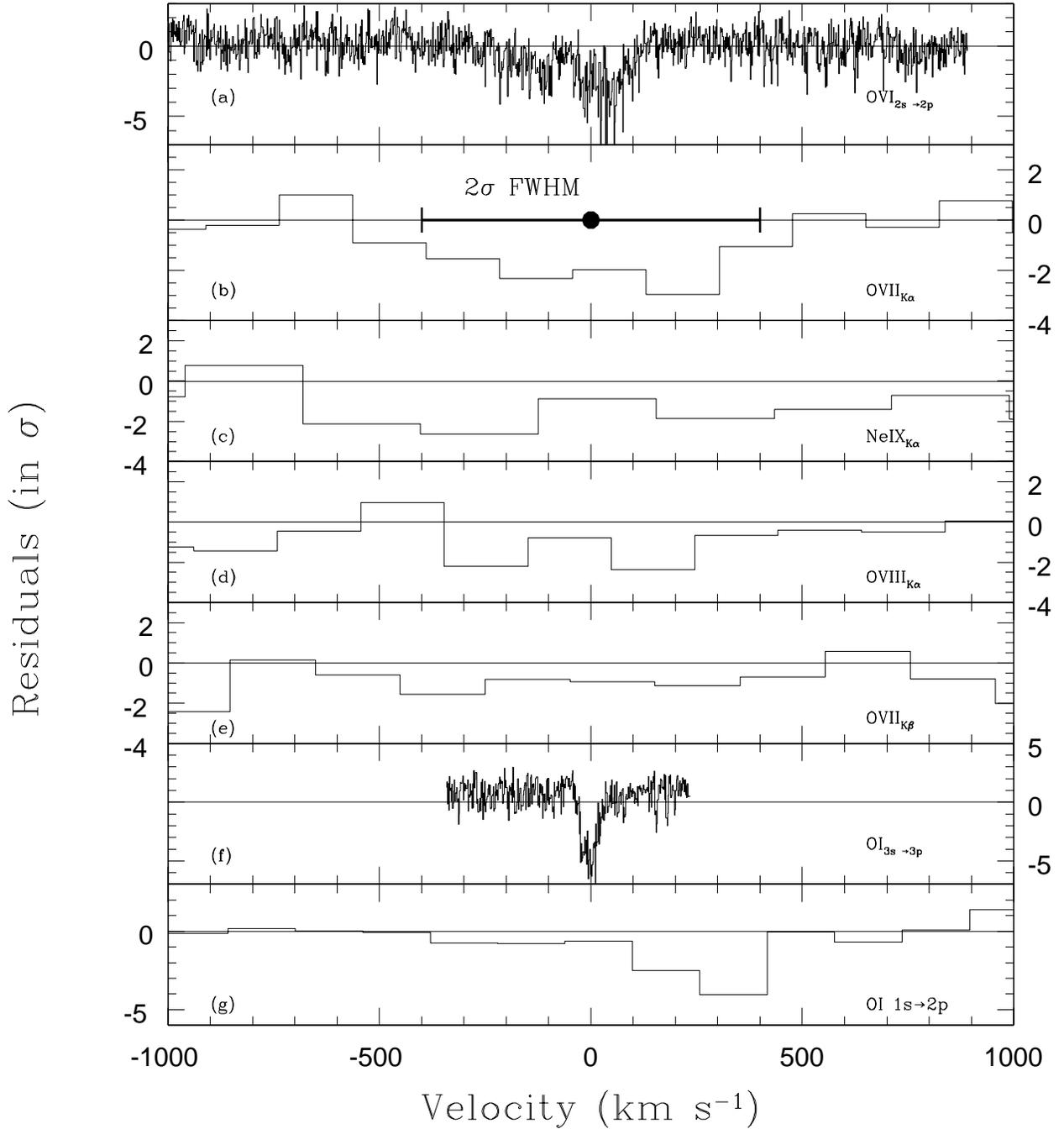} 
\vspace{0in}\caption[h]{\footnotesize Residuals, in velocity
space, of the relevant portions of the FUSE (a,f) and HRCS-LETG
(b,c,d,e,g) spectra of PKS~2155-304, to their best fitting
continuum models. The residuals are centered, from the top to the
bottom panels, on the rest-frame wavelengths of the resonant
transitions: OVI$_{2s\rightarrow 2p}$, OVII$_{K\alpha}$,
NeIX$_{K\alpha}$, OVIII$_{K\alpha}$, OVII$_{K\beta}$,
OI($\lambda=23.489$). Bin widths are as follow: 2 km s$^{-1}$
(FUSE: panel a,f), 190 km s$^{-1}$ ({\em Chandra}: panels b, d,
e), 160 km s$^{-1}$ ({\em Chandra}: panel g), and 280 km s$^{-1}$
({\em Chandra}: panel c).}
\end{figure}
%

To intercompare the dynamics of the gas producing the detected
absorption lines, figure 3 shows their residuals, in velocity
space, from the FUSE and {\em Chandra} spectra to their best
fitting continuum models. Zero velocities in this diagram
correspond to the rest frame wavelengths of the detected
transitions (Verner et al., 1996, Krause, 1994).  The spectral
resolution of the HRCS-LETG (R$\sim 450$, FWHM $\simeq$ 660 km
s$^{-1}$ at 20 \AA) does not resolve the high ionization X-ray
absorption lines (see Table 1).

The spectral resolution of the FUSE spectrometer ($R \sim 15000$,
FWHM $\simeq$ 20 km s$^{-1}$ at $\sim 1000$ \AA) is more than one
order of magnitude larger than that of the {\em Chandra}
HRCS-LETG, and is sufficient to resolve the OVI$_{2s\rightarrow
2p}$ absorption into at least two components (Figure 3a; Table
1).  The total width of the OVI$_{2s\rightarrow 2p}$ complex in
the FUSE spectrum is FWHM=264$\pm 35$ km s$^{-1}$, fully
consistent with the upper limit on the width of the high
ionization X-ray lines (Table 1).  We also note that both
position and width of the OI$_{3s\rightarrow
3p}$($\lambda=1039.23$) absorption line in the FUSE spectrum, are
clearly different from widths and positions of each of the two
OVI components in the same spectrum (Fig. 3a,f), but again fully
consistent with width and position of the OI$(\lambda=23.489)$
line in the {\em Chandra} spectrum (Fig. 3f,c, Table 1).

We then conclude that: (a) the high ionization lines, both in the
UV and X-rays are likely to be produced by gas with similar
dynamical properties, and broad range of dispersion velocities
(from the UV profile), while (b) the low ionization atomic OI
lines (both in FUSE and {\em Chandra}) are produced by a
slow-moving cloud of cold gas, with relatively low internal
dispersion velocity, and so differs dramatically in both physical
and dynamical properties from the high ionization gas.

\noindent
In the following we discuss only the highly ionized component.

\subsection{Curve of Growth Analysis}

Absorption lines equivalent widths (EWs), and their ratios, can
in principle provide invaluable information on the physical state
of the gas, and allow one to distinguish between quite different
scenarios.  However these diagnostics can only be applied to
unsaturated lines.  Fortunately the UV OVI$_{2s\rightarrow 2p}$
lines are clearly not saturated, since the ratio between the EWs
of the doublet ($\lambda_1 = 1031.9261$ \AA and $\lambda_2 =
1037.6167$ \AA), is fully consistent with the oscillator
strengths (OSs) ratio of 2 (Savage et al., 2000).  For the X-ray
lines of OVII, OVIII and NeIX, in principle, one could compare
the EWs of the corresponding K$\alpha$ and K$\beta$
lines. However, the errors on the EWs of the K$\beta$ lines from
these ions are too large.  An alternative approach uses the Curve
of Growth (CoG; e.g. Spitzer, 1978), computed for a given Doppler
parameter, to see whether each line falls on the linear branch of
the CoG.  We ran our model for resonant absorption (Nicastro,
Fiore and Matt, 1999) to produce CoGs for the OVI$_{2s\rightarrow
2p}(\lambda=1031.9261)$, the OVII$_{K\alpha}$, the
OVIII$_{K\alpha}$ doublet, and the NeIX$_{K\alpha}$.  We assume a
Doppler parameter of b=200 km s$^{-1}$, since this is the
observed $\sigma$ of the OVI$_{2s\rightarrow 2p}$ complex.  The
CoGs are plotted in Figure 4, along with the maximum possible
measured EWs (best fit + 2$\sigma$) of each line (EWs are plotted
in absolute value).  For the OVI$_{2s\rightarrow
2p}(\lambda=1031.9261)$ complex we considered the total EW from
the two best fit gaussians.  All the lines EWs fall well within
the linear branches of their CoGs. Hence the lines
are not saturated, and the resulting column densities, listed in
Table~1, should be reliable. 
\footnote{The partial covering and scattering effects that can
occur when the absorbing material is close to the source, as in
AGN ``warm absorbers'' and broad absorption lines (e.g. Arav et
al., 2001), do not apply in this case, since PKS~2155-304 has
z=0.116, while the absorbers are at their rest wavelengths, and so 
are physically well separated.}
. 

%
\begin{figure}
\plotone{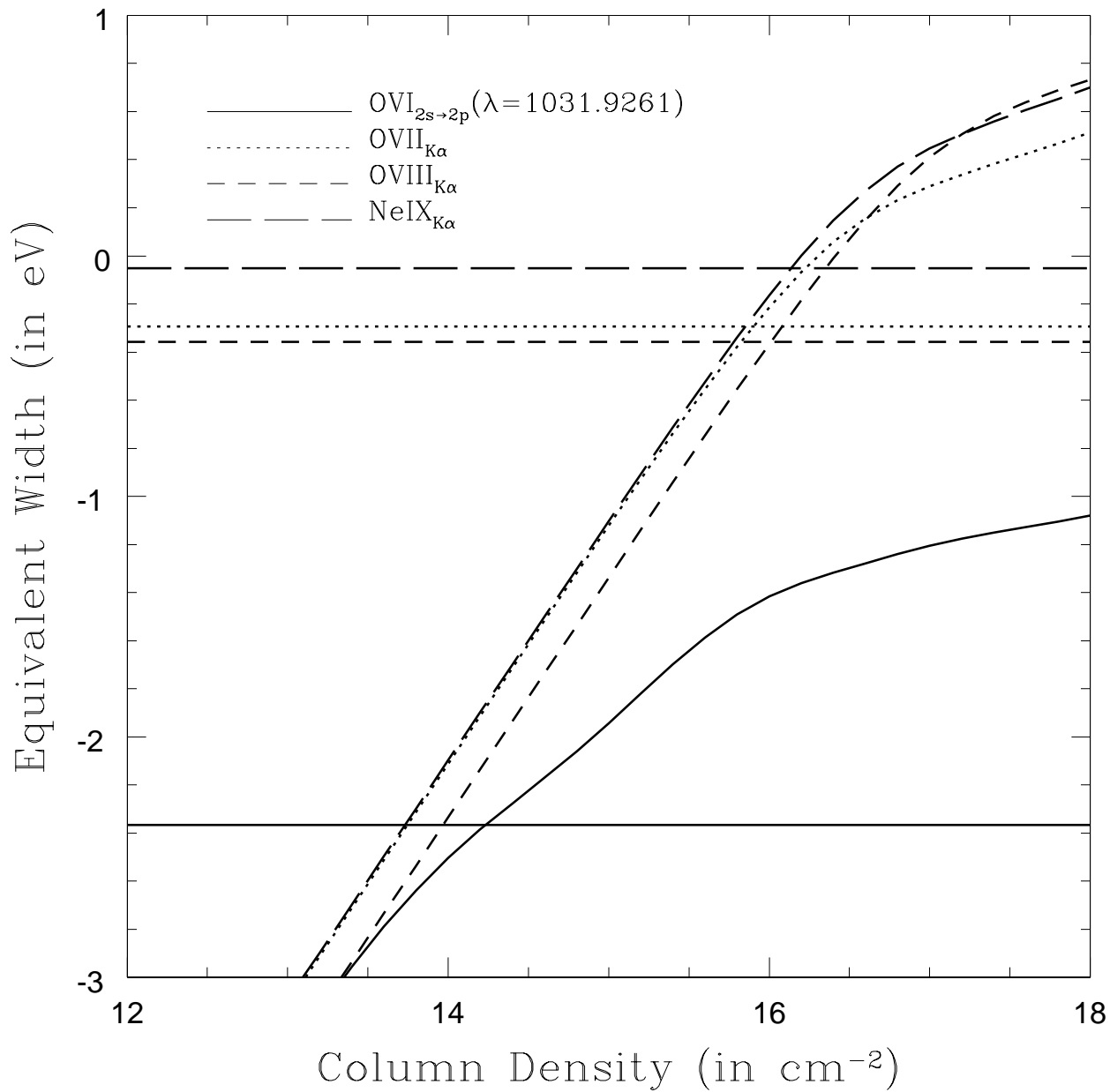} 
\vspace{0in}\caption[h]{\footnotesize COGs for the detected UV
and X-ray absorption lines: OVI$_{2s\rightarrow
2p}(\lambda=1031.9261)$, OVII$_{K\alpha}$, OVIII$_{K\alpha}$
(doublet), and NeIX$_{K\alpha}$.  A doppler parameter of b=200 km
s$^{-1}$ (the observed $\sigma$ of the OVI$_{2s\rightarrow 2p}$
complex) is assumed. Horizontal solid lines are the maximum
possible observed EWs (best fit + 2$\sigma$).}
\end{figure}
%

\section{Discussion}

The FUSE and {\em Chandra} spectra of PKS~2155-304 contain
absorption features from local intervening ionized gas that span
a broad ionization interval, from neutral (OI), through mildly
ionized (OVI, OVII), to highly ionized (OVII, OVIII and NeIX).
The neutral and the ionized components have very different
dynamical and physical properties and so must belong to two
different, spatially separated, systems.  The profile of the OVI
complex in the FUSE spectrum suggests that the ionized component
is also made up of at least two different absorbers: one
slow-moving, with relatively low internal dispersion velocity and
possibly associated with a cloud in the Galactic disk, and one
moving towards us with a velocity of about 150 km s$^{-1}$ ($\sim 
100$ km s$^{-1}$ transforming from the Heliocentric to the Galactic 
Standard of Rest frame
\footnote{http://nedwww.ipac.caltech.edu/forms/vel\_correction.html}
) and with a similar internal dispersion velocity 
and possibly associated either with an ionized high velocity cloud in the
Galactic halo, or with a WHIM filament.

In the next sections we demonstrate that neither pure collisional
ionization nor pure photoionization models can account for the
presence of strong OVII and NeIX absorption with the observed
intensity of the OVI lines, unless the three absorbers (one X-ray
and the two UV) are associated with three different Galactic
clouds with drastically different equilibrium conditions (with
temperatures differing by an order of magnitude), and so
locations.  Alternatively we show that if there is one highly
tenuous absorber, then the photoionization contribution by the
diffuse X-ray background becomes important and can change the
internal ion abundance distribution to account for all the UV and
X-ray observational constraints in a single absorbing medium.
These solutions however, imply path lengths of the order of several Mpc,
and so definitively locate the UV-X-ray absorber in the
intergalactic medium (IGM), beyond our Galaxy on a scale of the
Local Group, or somewhat bigger.


\subsection{The XRB Contribution}

For the photoionization contribution by the diffuse,
extragalactic X-ray background, we adopt an EUV to $\gamma$-ray
Spectral Energy Distribution [i.e. the specific flux
$F_{bkg}(E)$] given by a broken power law with an exponential
high energy cutoff (i.e.  $F_{bkg}(E) = \left[ K1
(E)^{-\alpha_{soft}} + K2 (E)^{-\alpha_{hard}} \right]
e^{-(E/E_f)}$, with $K1=0$ for $E>E_b$, $K2=0$ for $E<E_b$,
$\alpha_{soft} = 1.1$, $\alpha_{hard} = 0.4$, $E_{b} = 0.7$ keV,
and $E_{f} = 50$ keV; we also adopt a low energy cutoff at energies 
lower than the hydrogen ionization threshold $E_{HI}$).
We assume a normalization of 10~ph~s$^{-1}$ cm$^{-2}$ keV$^{-1}$
sr$^{-1}$, at 1 keV (Parmar et al., 1999, Boldt, 1987, and Fabian
\& Barcons, 1992).  Following the Hellsten et al. (1998)
parameterization for the redshift evolution of the X-ray
background, we can write the ratio between the diffuse background
ionizing photons density and the electron density of a filaments
at redshift z (i.e., the ionization parameter U) as:
\begin{equation} \label{uxrb}
U = [(1 + z)^3 / 4 \pi c] n_e^{-1} \int_{\nu_{HI}}^{+\infty} 
d\nu 4 \pi (F_{bkg}(\nu) / h\nu) \simeq 1.4 \times 10^{-7}
n_e^{-1}  (1 + z)^3 = 0.7 \delta^{-1} (\Omega_b h^2 /
0.02)^{-1},  
\end{equation} 

where $\nu_{HI}$ is the frequency of the H ionization threshold. 

In eq. \ref{uxrb} we defined the local IGM overdensity compared
with the mean density of the Universe, $\overline{n_e}$, as
$\delta = n_e / \overline{n_e}$, where $\overline{n_e} = (2
\times 10^{-7} (1 + z)^3 (\Omega_b h^2 / 0.02)$ cm$^{-3}$
(Hellsten et al., 1998).  Hence, for typical overdensities of
$\delta = 10$ (Hellsten et al., 1998, Dav\'e et al., 2000), we
expect, at any redshift, $U \sim 0.1 \delta_{10}^{-1}$, for
$\Omega_b h^{2} = 0.0125$ (Walker et al., 1991; h = 0.5). Given
the flat X-ray spectrum of the X-ray background, XRB
photoionization greatly modifies the ionic distribution of a
purely collisionally ionized gas at temperatures lower than logT
$\sim 6.4$, for electron densities of $n_e \ls 10^{-4}$ cm$^{-3}$
($\delta \ls 800$).  Photoionization is not important at typical
ISM densities of $10^{-2}$ to 1 cm$^{-3}$ (e.g. Spitzer, 1990;
but see \S 4.5 for alternative ionizing sources).

\subsection{Models}

We ran CLOUDY (vs. 90.04, Ferland, 1997) to build models for
temperatures (in K) from logT = $5.1 - 8$ and densities
$n_e=10^{-7}-1$ cm$^{-3}$.  Figure 5 shows the ratios between the
predicted ionic abundances of OVII/OVI (black curves), OVIII/OVI
(green curves), OVIII/OVII (red curves) and NeIX/OVII (blue
curves). Solid curves are for $n_e = 1$ cm$^{-3}$, while
long-dashed curves correspond to $n_e = 10^{-6}$ cm$^{-3}$
(i.e. $\delta \sim 10$). The thick sections of the lines delimit
the 2$\sigma$ allowed intervals given the measured relative ion
column densities derived from the corresponding EWs (see
Table~1), using the formula - valid only when the lines are not saturated - 
$EW(X^i) \propto A(X) n_{X^i}$ (Nicastro, Fiore, \& Matt, 1999), where 
$A(X)$ is the relative abundance of the element X, compared to H, and 
$n_{X^i}$ is the relative density of the ion $i$ of the element $X$. 
Figure 5 uses the total measured EW for the OVI$_{2s\rightarrow 2p}$ 
complex (i.e. narrow and broad components combined).

According to the above formula, the ratio between equivalent widths 
of two lines from two different ions of the same element, then, depends 
only on the ratio of the corresponding equivalent widths. Instead, for ions 
from different elements, the relative ion densities are proportional to the 
inverse of their relative metallicity (i.e. $n_{X^i}/n_{Y^j} \propto 
(EW(X^i)/EW(Y^j)) \times (A(X)/A(Y))^{-1}$). 
For the NeIX/OVII ratio (blue lines) we show allowed regions for a solar 
Ne/O ratio (magenta, dotted line, delimited by perpendicular tickmarks) and 
for 2.5 times solar ([Ne/O]$_{solar} = 0.158$, Grevesse \& Anders, 1989, 
Grevesse \& Noels, 1993. Blue, dashed line). 
This values are discussed in \S 4.3.

%
\begin{figure}
\plotone{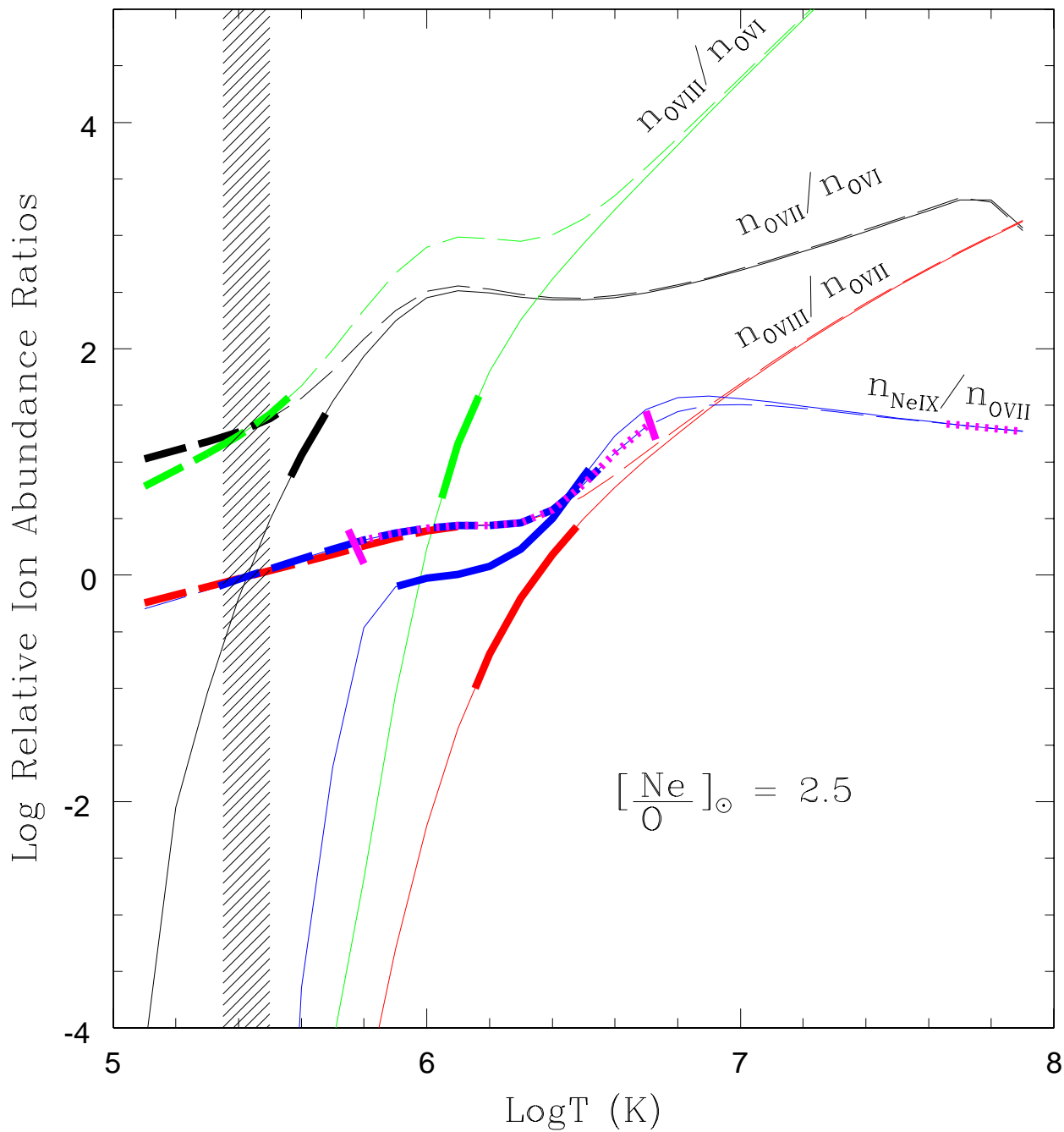} 
\vspace{0in}\caption[h]{\footnotesize Ratios between the
predicted ionic abundances of OVII/OVI (black curves), OVIII/OVI
(green curves), OVIII/OVII (red curves) and NeIX/OVII (blue
curves), for two electron densities: (a) $n_e = 1$ cm$^{-3}$
(solid curves), and (b) $n_e = 10^{-6}$ cm$^{-3}$ (long-dashed curves). 
Thick tracks on these curves
delimit the 2$\sigma$ allowed intervals for the measured relative
abundances ratios for the same ions, derived for the total
measured EW of the OVI$_{2s\rightarrow 2p}$ complex.  Ne/O
ratios of 1 (magenta, dotted line) and  2.5 (blue, dahed line) times solar 
are plotted.}
\end{figure}
%

\subsection{Galactic versus Extragalactic Origin}

Let us consider first the Galactic ``high-density'' solutions,
for which the gas is in pure collisional ionization equilibrium.
The ratio of OVII$_{K\alpha}$ to the entire OVI$_{2s\rightarrow
2p}$ UV complex (thick solid black curve) sets a quite stringent
upper limit of logT $\ls 5.7$ (T in K) on the allowed temperature
of the OVI-OVII absorber, if both the lines are produced by the
same material.  However, the OVIII and NeIX ions are clearly too
highly populated to be consistent with the intensities of the OVI
and OVII lines, requiring logT$> 6.1$, and so no common solution
exists for all the lines. We find that this is true for any $n_e
\gs 10^{-2}$ cm$^{-3}$.

Let us now consider the extreme extragalactic ``low-density''
($n_e = 10^{-6}$cm$^{-3}$) solution (Fig. 5, thick long-dashed
curves) where photoionization by the diffuse XRB overionizes the
medium.  These conditions allow the oxygen OVI, OVII, and OVIII
ions to have a common temperature (shaded region in Fig. 5) for
the relatively low temperature range $5.35 \le$ logT $\le 5.50$.
A consistent solution for even the oxygen lines, requires the
absorber to be highly tenuous ($n_e \ls 5 \times 10^{-6}$
cm$^{-3}$). A solar Ne/O ratio (magenta, dotted intervals) is not consistent 
with these conditions, but a Ne/O ratio of $\sim 2.5$ times solar (blue, 
dashed interval) provides consistency for all the ions (see \S 4.7).
Increasing the Ne/O metallicity ratio decreases the 
NeIX/OVII relative density ratio estimated from the corresponding measured 
EW ratio - see \S 4.2 - and so lets the thick intervals in Figure 5 shift 
toward the left on the theoretical $n_{NeIX}/n_{OVII}$ curve, down to the 
region where a common solution for all the ions can be found. 

For a given temperature, we can derive the equivalent H column
density of this gas. Assuming Ne/O = 2.5 and 
logT = 5.45 gives $N_H \simeq 1.5 \times 10^{19} [O/H]_{\odot}^{-1}$ 
cm$^{-2}$.
This translates, for a constant electron density of $n_e =
10^{-6}$ cm$^{-3}$, to a linear size of the absorbing cloud of
gas of $4.9 [O/H]_{\odot}^{-1}$ Mpc, consistent with a filament
with an overdensity of $\sim 10$ located in the intergalactic
medium near to our Galaxy.  For O abundances of 0.3
$\times Solar$ (as found in clusters of galaxies, Sarazin, 1988) the 
size would increase to $\sim 15$
Mpc ($\sim 2$ \% of the distance to PKS~2155-304, and $\sim 25$
\% of the distance to the first group of galaxies close to our
line of sight to the blazar: Penton Stocke, \& Shull, 2000).

\subsection{Multi-Component Solutions}

The above discussion is based on the hypothesis that both the
broad and narrow components of the OVI absorption complex are
produced by a single, structured, absorber that is also producing
the X-ray absorption.  Here we explore alternative
multi-component solutions.

Let us suppose that only one of the two OVI$_{2s\rightarrow 2p}$
absorption lines in the UV spectrum, is associated with the X-ray
absorber. The broader OVI line contributes about 1/3 to the
total observed equivalent width from the OVI system, while the
remaining 2/3 are provided by the narrower line (see Tab. 1).
Based on the strengths of the OVI$_{2s\rightarrow 2p}$ lines and
on the above discussion, it is then natural to associate the
broader UV component with the X-ray absorber.  In this case the
narrow and more intense OVI component would be produced by a
Galactic cloud of moderately ionized gas in collisional
equilibrium.  This ``narrow'' line gas must have a temperature of
logT$^{narrow} \gs 5.0$ if the size of the cloud is not to exceed
1 kpc, the size of a big halo cloud (e.g. Spitzer, 1990), given
the measured EW and assuming a maximum Galactic ISM density of
$\sim 1$ cm$^{-3}$.  The upper limit on the temperature, instead
is set by the observed EW(OVI$_{2s\rightarrow
2p}^{narrow}$/EW(OVII$_{K\alpha}$) ratio, and this gives
logT$^{narrow} << 5.6$.

The second absorber should instead produce the broader and weaker
OVI line, and all the X-ray lines. We tested this hypothesis by
building a diagram similar to that in Fig. 5, but considering a
variable fraction (from 0 to 1) of the EW of the broad
OVI$_{2s\rightarrow 2p}$ line (Figure 6).
\noindent 
We find that equilibrium, high density solutions exist only if
a fraction of 5 to 25 \% (at 2$\sigma$) of the broad OVI line is produced 
by the same gas producing the X-ray absorption. These solutions are limited 
to a range of temperatures around logT=6.4, whose width depends on the 
relative [Ne/O] abundance: $\Delta(logT)=0.16-0.32$ for [Ne/O]$_{\odot}=1-2$ 
(dashed regions in Fig. 6). For logT = 6.4 and 
[Ne/O]$_{\odot}=1$, we find $N_H = 2.2 \times 10^{19} [O/H]_{\odot}^{-1}$, 
and so a linear size of $D = 7 [O/H]_{\odot}^{-1}$ pc, for $n_e = 1$ 
cm$^{-3}$.

%
\begin{figure}
\plotone{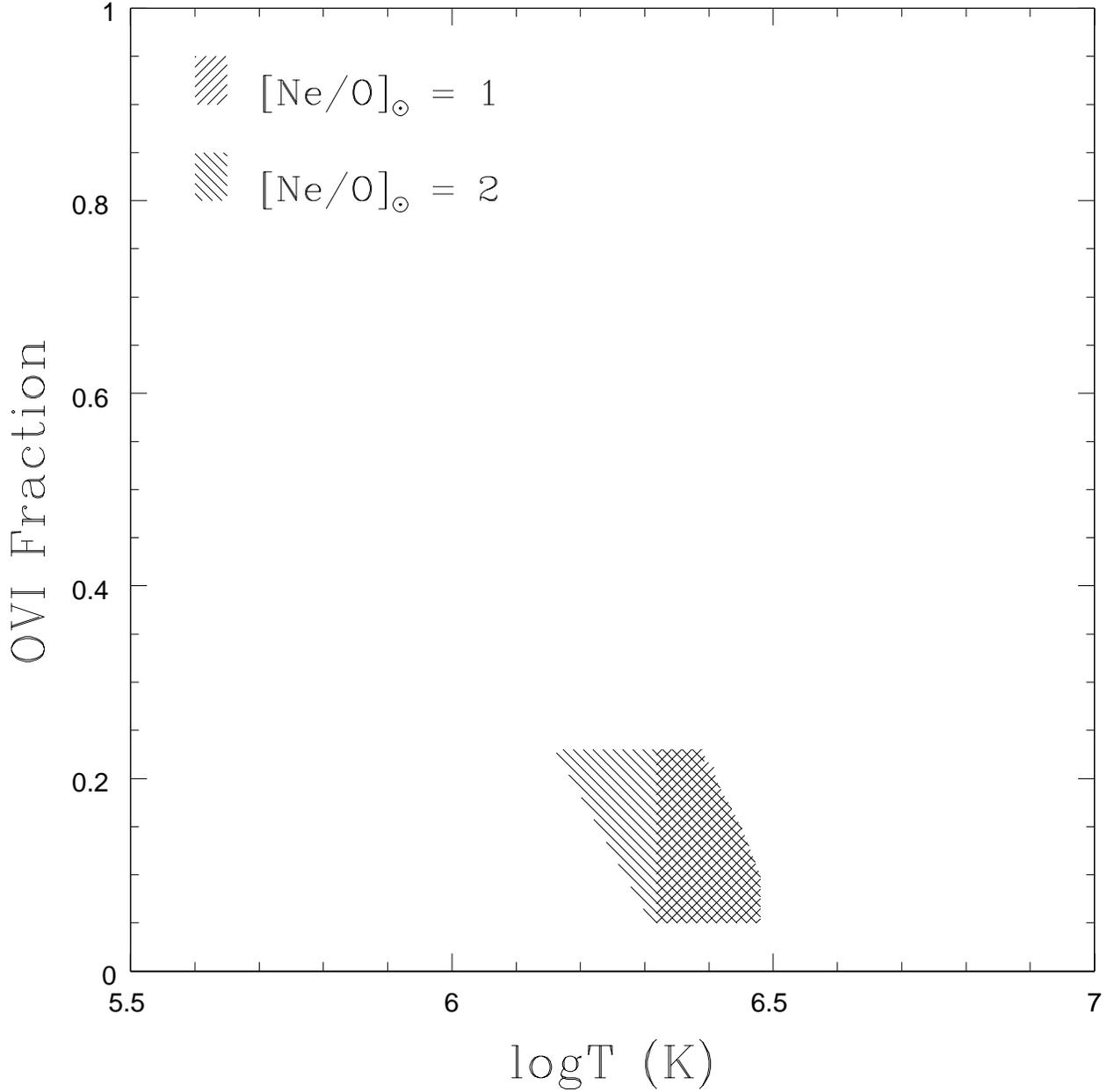} 
\vspace{0in}\caption[h]{\footnotesize Fraction of broad OVI line (f(OVI)) 
produced by the same gas producing the X-ray absorption, as a function of 
gas temperature. The diagram shows the two regions [for [Ne/O]$_{\odot} =$ 
1 (red), and [Ne/O]$_{\odot} =$ 2, (blue)], in the f(OVI)-logT space, for 
which common UV-Xray Galactic solutions do exist: at most 25 \% of the 
broad OVI$_{2s\rightarrow 2p}$ can come from a Galactic cloud of X-ray 
absorbing gas}
\end{figure}
%

Since only 25 \% of the broad OVI$_{2s\rightarrow 2p}$ can come
from X-ray absorbing gas, a third {\em ad hoc} Galactic absorber 
similar in dynamical properties to the absorber imprinting the
broad OVI$_{2s\rightarrow 2p}$ line, but with temperature differing 
by at least an order of magnitude from the temperature of the UV-X-ray 
absorber, is required to explain the remaining broad OVI absorption. 
Alternatively, a highly structured and, again, {\em ad hoc} multi-temperature 
plasma, with at least two zones separated by a sharp front, and with the same 
kinematics, would be needed to account for all the observed UV and X-ray 
lines. 
We can broaden the range of allowed temperature and increase the fraction 
of OVI absorption produced by the X-ray absorber, by allowing both the 
Ne/O ratio to be supersolar, and the gas density to be lower than $\sim
10^{-4}$ cm$^{-3}$. However this converges to the extragalactic
solution (\S 5.2). We find that an acceptable solution with a
minimum size is obtained for logT = 5.7, $n_e = 6 \times 10^{-6}$
(i.e. $\delta \sim 60$), and Ne/O = 2, which gives $N_H = 6 
\times 10^{19} [O/H]_{0.3\times\odot}^{-1}$ cm$^{-3}$ and $D = 3 
[O/H]_{0.3\times\odot}^{-1}$ Mpc.

\subsection{Alternative Photoionizing Sources} 

As we have demonstrated (\S 4.1, 4.3) X-ray photoionization is
needed to modify the ion relative abundance distribution from the
pure collisionally ionized case (with logT = 5.3-5.7), to allow
for the simultaneous presence of OVI, OVII and OVIII, and also
NeIX. In order for the photoionization contribution to play a
significant role, the ionization parameter 
has to be of the order of $U \sim 0.1$ (\S 4.1). 
Since the ionization parameter. $U$, is simply the ratio between the
number density of ionizing photons and the electron density of
the gas, $U$ increases as either the electron density in the gas decreases (as
in the IGM case) or the distance between the source of ionizing
photons and the gas cloud decreases. So high values of U might also be
reached in gas with typical ISM densities (i.e. $n_e \sim
10^{-2}-1$ cm$^{-3}$) illuminated by other, discrete and locally
much brighter hard X-ray sources than the XRB. In this section we explore
this second possibility.

\subsubsection{Isotropic Nearby Source}
Based on the needed value of $U \sim 0.1$ we can put constraints on 
the brightness of the ionizing source as seen by us, and on the angular 
separation between the source and the gas cloud. 
Let us consider an intervening cloud of gas located at a distance $r$ from 
us, along a given line of sight, and an ionizing source, located at a distance 
$D$ from us. Let $d$ be the distance between the ionizing source and 
the gas cloud. We can write: (a) $L = 4 \pi D^2 F_{obs}$; (b) $L = 4 \pi 
d^2 F_{ion}$; and (c) $r tg\theta = d$. In the above equations, 
$L$ is the source luminosity, and $F_{obs}$ and $F_{ion}$ are the source 
fluxes as seen by us and the gas cloud respectively. These two fluxes are 
related to each other through the source-gas angular separation $\theta$: 
$F_{ion} = (sin\theta)^{-2} F_{obs}$. The ionization parameter U at the 
illuminated face of the cloud can then be written as a function of the 
source flux as seen by us: $U = q n_e^{-1} c^{-1} (sin\theta)^{-2}$, 
where $q = \int_{E_{HI}}^{+\infty} dE E^{-1} f_{obs}(E)$ is the total 
flux of ionizing photons ($F_{obs} = \int dE f_{obs}(E)$). 
To maximize the contribution from hard X-ray 
ionizing photons we use as the ionizing continuum a cutoff power law with 
$\Gamma = 1.5$ from the H ionization threshold up to 100 keV, and $\Gamma =
4$ from 100 keV up to infinity. This gives a total flux of ionizing 
photons of $q \simeq 1.3 \times 10^9 F^{2-10}_{obs}$ ph s$^{-1}$ (where 
$F^{2-10}_{obs}$ is the observed flux between 2 and 10 keV, and we 
assumed $F_{obs} = 3 \times F^{2-10}_{obs}$). 
Our value $U \sim 0.1$ then gives: 
$U \simeq 4 \times 10^{-2} n_e^{-1} (F^{2-10}_{obs})_{CGS} (sin^{-2}
\theta) \sim 0.1$. For $\theta = 30$ arcmin and gas densities in the ISM 
range $n_e = 10^{-2}-1$ cm$^{-3}$, then, the 2-10 keV observed flux must be as 
large as $F^{2-10}_{obs} \sim 2 \times 10^{-6}-10^{-4}$ erg s$^{-1}$ 
cm$^{-2}$ ($ \sim 10^2-10^4$ Crab), to efficiently ionize the gas. For a 
30 times smaller separation, this fluxes reduce to $2 \times 10^{-9}-10^{-7}$ 
erg s$^{-1}$ cm$^{-2}$ ($ \sim 0.1-10$ Crab). 
No such source is currently seen. 

\noindent
For halo distances of 30~kpc the implied 2-10 keV luminosities are 
$L^{2-10}_{obs}(\theta=30') = 2 \times (10^{41}-10^{43})$ erg s$^{-1}$ 
($d = 260$ pc), far too large for any known stellar-size Galactic source, 
and $L^{2-10}_{obs}(\theta=1') = 2 \times (10^{38}-10^{40})$ erg s$^{-1}$ 
($d = 8$ pc), just about the maximum observed luminosity for the most 
luminous known Ultra-Luminous-X-ray  sources (ULXs) or beamed microquasars. 
For the much closer distances of Galactic disk objects, say, 10 pc, these 
luminosities reduce to more normal values for X-ray binaries (i.e. 
$L^{2-10}_{obs}(\theta=30') = 2 \times (10^{34}-10^{36})$ erg s$^{-1}$, and 
$L^{2-10}_{obs}(\theta=1') = 2 \times (10^{31}-10^{33})$ erg s$^{-1}$) 
but the physical separations between the source and the gas cloud become 
very small: $d = 0.09$ pc and $d = 0.003$ pc respectively. 
So, either very luminous sources, or very close source-to-gas systems 
are needed to efficently ionize a tenuous gas cloud in our Galaxy to the 
observed degree. 

We searched the ROSAT-WGACAT (White, Giommi, \& Angelini, 1994)
for bright X-ray sources in a 30 arcmin radius cylinder, centered
on the line of sight to PKS~2155-304. We found 33 unidentified
sources plus an F0 star with ROSAT-PSPC count rates between
0.0019 and 0.024 ct s$^{-1}$, and angular distances from the line
of sight from $\theta = 2.1$ to $\theta = 20.4$ arcmin.  The
closest of these sources to the PKS~2155-304 line of sight has a
count rate of 0.0073 (source A: 1WGA~J2158.8-3011), while the brightest 
(0.025 ct s$^{-1}$) is at a distance of 15.1 arcmin (source B: 
1WGA~J2158.2-3000). We used PIMMS
\footnote{http://legacy.gsfc.nasa.gov/Tools/w3pimms.html}
to convert these count rates, into flux.  To maximize the
possible contribution at energies E$> 1$ keV, we assumed a hard
X-ray continuum (a powerlaw with $\Gamma = 1.5$) heavily absorbed
by a column of $N_H = 10^{22}$ cm$^{-2}$ intrinsic neutral gas.
This gave 2-10 keV fluxes of $10^{-12}$ erg s$^{-1}$ cm$^{-2}$ ($5 
\times 10^{-5}$ Crab), for source A, and $3.6 \times 10^{-12}$ erg s$^{-1}$ 
cm$^{-2}$ ($1.8 \times 10^{-4}$ Crab) for source B, at least (i.e. 
using $n_e = 10^{-2}$ cm$^{-3}$) 4 and 5 orders of magnitude, respectively, 
lower than the values required to photoionize the gas at these 
angular separations. 

\subsubsection{Beamed Nearby Source}
Let us consider a microquasar highly beamed into the direction of the gas 
cloud,and whose line of sight to us is somewhat perpendicular to the beam 
direction.
The probability for a microquasar to be beamed in the 
direction of the gas cloud, and located within a 20-130 pc radius sphere 
from the gas cloud, depends on the distribution of the parent poulation, 
that, for microquasars, is largely unknown. 
We can compute here the minimum beaming factor needed, in the above 
optimal geometrical configuration, for source A or B to efficiently ionize 
a cloud of gas along the line os sight to PKS~2155-304. 
The luminosity amplification due to beaming, in the direction of the jet, is
given by: $L_{beamed} = L_{intr} \times (2 \gamma)^{p}$.  The
observed luminosity along a perpendicular direction is insted
de-boosted by the factor: $L_{obs} = L_{intr} \times \gamma^{-p}$
(Urry \& Shafer, 1984). In the above equations, $L_{intr}$ is the source
intrinsic luminosity, $\gamma$ is the Lorentzian factor, and $p
\simeq 4$ (Urry \& Shafer, 1984). Solving for the intrinsic luminosity, and 
using as maximum value for the beamed luminosity of a microquasar $L_{beam} 
= 5 \times 10^{40}$ erg s$^{-1}$ and the observed value of $L_{obs} = 4 \times 
10^{35}$ erg s$^{-1}$ (for source B, at 30 kpc), we
obtain a minimum relativistic factor of $\gamma \ge 3$, to
compare with the estimated upper limit of $\gamma = 3$ for the
microquasar GRS~1915+105 (Mirabel \& Rodriguez, 1999). 
Any deviation from the above
ideal geometry, would give a much larger Lorentz factor.

\subsubsection{Historical Nearby Source}
Low density gas has long recombination times.  A source may
thus have been much more luminous in the past and ionized the gas
then, from which the gas has not yet recovered.  Let us first suppose
that the most luminous X-ray source we are currently seeing in the
ROSAT archive is a binary system that went through a ULX phase
(King et al., 2001) in the past (from the calculation in \S 4.5.1, for 
source A and B we need L(outburst)/L(now)$> 10^4-10^5$, respectively 
\footnote{Such outbursts have been observed in X-ray novae between quiescent 
and active phases, e.g. Narayan, Garcia, \& McClintock, 2001.}
). 
The recombination time for the OVII-OVIII ions, in gas with
electron density of $n_e = 1$ cm$^{-3}$ is of about $t_{rec} \sim
10^5$ years (Nicastro et al., 1999). Typical ULX phases are
thought to last for $\sim 10^5$ years, and in our Galaxy the
total estimated rate of ULXs is of 1 every $10^{5}$ years (King et al., 
2001). 
So, allowing for a maximum delay of $t_{delay} = t_{rec} = 10^5$
years, we may expect N = 2 ULXs in all the Galaxy over a period
of $\Delta t = t_{phase} + t_{delay} = 2 \times 10^{5}$
years. Assuming, for the halo, a 30 kpc radius sphere, this gives a ULX
density, over the same period, of n = $2 / ((4/3) \pi \times (30)^3
\hbox{kpc}) \simeq 1.8 \times 10^{-5}$ kpc$^{-3}$. 
So, the maximum number of ULXs expected over a period $\Delta t$
in a random $R_{max} = 20-130$ pc radius sphere in our galaxy (the 
physical distances between sources A and B and the line of sight 
to PKS~2155-304, if both sources are put at 30 kpc from us, in the 
outskirts of the Galactic halo) is of $N_{max} = 6 \times 10^{-10}- 2 \times 
10^{-7}$, making prior ionization by a powerful transient source highly 
unlikely, albeit with {\em a posteriori} statistics.

\subsubsection{Historical AGN at Galactic Center}
A final possibility is that in the past the center of our Galaxy
went through an AGN phase, producing far higher luminosities 
and photoionizing any low-density ($n_e \sim
10^{-2}$ cm$^{-3}$) clouds in the Galactic halo. At these
densities we have $t_{rec}(OVII,OVIII) \sim 10^{7}$ years.  Let
us then suppose that less than $10^7$ years ago the $2 \times 10^6$
$M_{\odot}$ black hole in the center of our Galaxy (Chakrabarty \& Saha, 2001) 
was emitting at its Eddington luminosity of $2 \times 10^{44}$ erg~s$^{-1}$. 
Assuming a typical Spectral Energy Distribution (SED) for AGN (Mathews \&
Ferland, 1987) this would produce an ionization parameter of
$U_{AGN} = 2.4 \times 10^{45} \times D^{-2}$ at the surface of
$n_e = 10^{-2}$ cm$^{-2}$ halo clouds located at a distance D
from the central AGN. To produce the observed line ratios, the
photoionization contribution must be such that $U_{AGN} \sim 0.1$
at the illuminated face of such a cloud of gas.  This gives a
distance for the gas cloud of $D \sim 36$ kpc, consistent with the outermost 
size of the Galactic halo. 
This solution requires that all halo clouds would show UV and
X-ray absorption with similar physical properties as those
observed in PKS~2155-304.
Using the much lower luminosities estimated by ASCA for the central AGN in 
our Galaxy about $10^4$ years ago (i.e. $L \sim 10^{41}-10^{42}$ erg 
s$^{-1}$, Koyama et al., 1996), gives distances of 1-5 kpc, well within the 
Galaxy's halo, and therefore not consistent with the the galactic coordinates 
of PKS~2155-304 (given our distance of about 10 kpc from the Galactic center). 

\subsection{Properties of the WHIM Gas}

From our analysis of the combined UV and X-ray data, we conclude
that the X-ray absorption and at least the broad
OVI$_{2s\rightarrow 2p}$ are likely to be produced in tenuous
extragalactic medium, with $n_e \sim 6 \times 10^{-6}$ cm$^{-3}$.
From the position of the OVI$_{2s\rightarrow 2p}^{broad}$ and
assuming that the feature lies within a few Mpc from our Galaxy,
the gas is falling toward our Galaxy with a bulk velocity of $\sim 100$ 
km s$^{-1}$ (in the Galactic Standard of Rest frame). 
The feature has a dispersion velocity ($\sim 300$ km~s$^{-1}$)
comparable with the bulk velocity, and a temperature of logT$\sim 5.7$. 
The implied overdensity is $\delta \sim 60$ and the total linear size of 
the feature along our line of sight is about $ 3 \delta_{60}^{-1} 
[O/H]_{0.3\odot}^{-1}$ Mpc. 

The Ne/O abundance can be reduced to Solar if we allow for
inhomogeneities in the flow, since inhomogeneities increase the
upper limits on both the temperature and the density, and so also
decrease the lower limit on the linear size of the feature (from
3 $[O/H]_{0.3\times\odot}^{-1}$ Mpc to $few \times 10^2
[O/H]_{0.3\times\odot}^{-1}$ kpc). Such inhomogeneities may arise near our
Galaxy halo, where the gas may mix with denser medium, cooling and producing
part of the observed OVI$_{2s\rightarrow 2p}^{broad}$ absorption.

\medskip
The above overdensities and linear sizes are consistent with
those predicted for the WHIM by hydrodynamical simulations for
the formation of structure in the local Universe (e.g. Hellsten,
1998), and in particular in our Local Group (Kravtsov, Klypin,
and Hoffman, 2001).  
According to these simulations, the local IGM is concentrated in
filaments with overdensities of $\delta \sim 5-100$ connecting
already virialized structured (galaxies and clusters of
galaxies), that were shock-heated during the collapse of the
density perturbations.  The average sizes of these perturbations
are of the order of 30-50 Mpc in the local Universe, but the
total extent of the density peaks are usually not larger than a
few Mpc. The velocities of the matter in these filaments, should
be of the order of the proper motions of the virialized
structures that they connect, $\sim 100-200$ km s$^{-1}$.  We
note that these velocities are of the same order as those
observed on the blue side of the structured and complex
OVI$_{2s\rightarrow 2p}$ profile in the UV spectrum of
PKS~2155-304.

We also note that both the dynamics of this gas (as derived from
the OVI line) and its physical state are very similar to those
found (using UV absorption lines only) for the intervening
ionized absorber along the line of sight to the quasar H~1821+643
(Tripp et al., 2001): internal dispersion velocities of the order
of $\sim 150$ km s$^{-1}$ (FWHM of the broad Ly$\alpha$ line),
temperature in the range $5.3 <$ logT $< 5.6$ (assuming pure
collisional ionization), and pathlength of about 3 Mpc (using
$H_0 = 50$ km s$^{-1}$ Mpc$^{-1}$). We predict that the planned
high resolution and high quality X-ray spectra of this object,
should then show OVII$_{K\alpha}$ absorption with an intensity
similar to that found here.

\subsection{The Ne/O problem: Dust Depletion}

In virtually all the cases discussed, the Ne/O ratio in the
absorber along the line of sight towards PKS~2155-304 is required
to be higher than solar. 
There are various ways to reconcile with
this problem: (1) Type II supernovae give rise to Ne/O ratios
much higher than Type I supernovae, compared to Si (Tsuro et al.,
1997 and references there in); (2) O can be depleted on to dust
grains, but noble gas like Ne would not. If superwinds from
galaxies are responsible for the enrichment of the IGM, then in
order to produce high gaseous Ne/O ratios compared to solar, they
must be able to produce a considerable amount of dust before
leaving the high density galaxy environment.  (3) In the modeling
above, we have assumed a simple parameterization of ionizing
background. The true shape of the background is clearly more
complex, and a harder background would result in a large Ne/O
ratio. Any or all of these factors might be playing a role in the
observed value of [Ne/O].

It is important to notice that dust can easily survive at the kinetic 
temperatures (i.e. $few \times 10^5$ K) and densities (i.e. $n_e \sim 
few \times 10^{-6}$ cm$^{-3}$) of such a warm IGM filament. 
In such regime, the sputtering lifetime is inversely proportional to the
gas density, and largely independent from the dust grain composition and
gas temperature (Draine \& Salpeter, 1979a,b). The survival time of a small
interstellar grain of size $\sim 0.01$~$\mu$m will be larger than
$10^{10}$~yr in the IGM filament, and even larger for grains of larger
size.
We estimated the reddening expected along the line of sight
to PKS~2155-304, based on (a) the observed equivalent H column density,
(b) the typical size of a silicate dust grain, and (c) a typical ISM
gas/dust ratio. In this condition the estimated E(B-V) is of about $2
\times 10^{-2}$ mag, fully consistent with the observations (Schlegel et 
al., 1998). 

Finally we note that [Ne/O] overabundances of $\sim 2-3$, compared to 
solar, have already been observed in several astrophysical environments 
(e.g. Paerels et al., 2001, Brinkman et al., 2001), strongly suggesting 
that the non-O-depleted values of this quantity, measured in stelar 
atmospheres, are not representative of the average gaseous [Ne/O] ratio 
in the Universe. 

\section{Conclusions}

In this paper we report the first discovery of the ``X-ray
Forest'' of absorption lines produced by highly ionized
intergalactic medium.  High significance X-ray absorption lines
of OVII, OVIII and NeIX are detected along the line of sight to
the bright blazar PKS~2155-304, and are associated with a known
OVI UV absorber.  We demonstrate that the dynamical properties of
the X-ray and the UV absorber are fully consistent with each
other, and that a reasonable common and self-consistent physical
solution can be found only if photoionization and collisional 
ionization both contribute to the ionization of the absorbing gas. 
This requires electron densities of about 
$6 \times 10^{-6}$ cm$^{-3}$ for the diffuse X-ray background to be 
significant. This low denisty requires a linear size,
along the line of sight, of the order of $3 [O/H]_{0.3\times\odot}^{-1}$
Mpc.  This clearly locates the absorber outside our Galaxy in
intergalactic space. We demonstrate that both the dynamical and
physical properties of such an absorber are remarkably consistent
with those predicted for the low redshift warm phase of the IGM
as predicted by hydrodynamical simulations for the formation of
structures in the Universe.  Finally we find that solutions with
a Ne/O ratio of about 2.5 times solar are favored, suggesting
type-II Supernova enrichment, or the presence of dust in the IGM.

\begin{center}
{\bf Aknowledgements}
\end{center}
This work has been partly supported by the NASA grant ADP
NAG-5-10814 (FN), the {\em Chandra} grant DDO-1005X FN and AF),
CXC grant NAS8-39073 (FN and AF) and the NASA grant LTSA
NAG-5-8913 (SM).



\end{document}